\documentstyle[12pt]{article}

\newcommand{\be}{\begin{equation}}
\newcommand{\ee}{\end{equation}}
\newcommand{\ba}{\begin{array}}
\newcommand{\ea}{\end{array}}
\newcommand{\bea}{\begin{eqnarray}}
\newcommand{\eea}{\end{eqnarray}}
\newcommand{\bt}{\begin{tabular}}
\newcommand{\et}{\end{tabular}}

\begin{document}

\title{SOME GENERAL PROBLEMS IN QUANTUM GRAVITY II: THE THREE
DIMENSIONAL CASE}

\author{Enrique \'Alvarez\\
\it{Departamento de F\'{\i}sica Te\'orica C-XI}\\
\it{Universidad Aut\'onoma de Madrid}\\
\it{28049 Madrid, Spain}}

\maketitle

\begin{abstract}
The general problems of three-dimensional quantum gravity  are
recatitulated here, putting the emphasis on the mathematical problems
of defining the measure of the path integral over all three-dimensional
metrics.This work should be viewed as an extension of a preceding one on
the four dimensional case (\cite{kn:eav5}), where also some general
ideas are discussed in detail. We finally put forward some suggestions
on the lines one could expect further progress in the field.

\end{abstract}

\newpage

\pagestyle{plain}

\baselineskip20pt

\newpage
\section{Introduction}
We have grown used to the notion of "euclidean quantum gravity" (cf.,
for example, ref. \cite{kn:eav}), thanks, among others, to the work of
Hawking and collaborators.
\par
It should nevertheless be stressed that this path integral stands apart
from other, superficially similar, expressions one could write for other
quantum fields,in the sense that  it is not a representation of a known
unitary evolution of states in a Hilbert space.
\par
It seems hardly arguable, however, that the problems encountered while
treating to make sense of the euclidean path integral, being of a very
general character, will also reappear in one guise or another in any
physical formulation of the problem.
\par
The purpose of the present work is to present some general observations
of the main characteristics of a sum over three-dimensional riemannian
geometries.
Before doing that, however, we review in the first paragraph Witten's
solution to three-dimensional quantum gravity as a Chern-Simons theory (
cf.
\cite{kn:witten1}\cite{kn:witten2}\cite{kn:witten3}\cite{kn:witten4}),
and we discuss in what sense it can be
considered as a complete solution of our problem.
In the second paragraph we explain some mathematical details on the
problem of the homeomorphic equivalence in $d=3$, in order to determine
in what sense gauge fixing is possible in the sum over all topologies.
In the fourth paragraph we dwell upon Thurston's geometrization
programme, an attempt at characterizing all three-dimensionsl manifolds
in geometrical terms.
\footnote{We will almost always reduce ourselves to compact, closed,
oriented three-dimensional manifolds; that is, the simplest situation}
In the fifth paragraph, we expound some recently discovered new
invariants, mostly related to knots, whose relation to Thurston's
viewpoint is still mostly unclear.
Finally, in the sixth and last paragraph, we conclude on a very
speculative note, trying to determine the main characteristics
of a well-defined measure, whose determination, if possible at all,
will require some deep mathematical research. as well as great doses of
physical intuition.

\section{Three-dimensional gravity as a topological Chern-Simons theory}
E. Witten (\cite{kn:witten1}, following earlier suggestions by Achucarro
and Townsend (\cite{kn:achucarro}), was the first to work out the
consequences of the point of view consisting in interpreting the spin
connection  $\omega$ as a gauge field for the Lorentz group, $SO(3,1)$,
and the vierbein as another gauge field corresponding to the translation
group, $T$.
\par
In first quantized formalism, the Einstein -Hilbert action for a
three-dimensional space-time manifold, diffeomorphic to $M = \Sigma
\times R$, where $\Sigma$ is a two-dimensional Riemann surface, can be
written as:
\be
S_{EH}=1/2 \int \epsilon^{ijk} \epsilon_{abc} e_{i}^{a}
(\partial_{j}\omega_{k}^{bc}-\partial_{k}\omega_{j}^{bc}
+[\omega_{j},\omega_{k}]^{bc})
\ee
It is not difficult to see that the preceding action is the Chern-Simons
action for the gauge field
\be
A_i = e_i^{a} P_a + \omega_i^{a} J_a
\ee
The gauge symmetries of the system are easily seen to be:
\be
\delta e_i^{a} = -\partial_i \rho^{a} - \epsilon^{abc} e_{ib} \tau_c -
\epsilon^{abc} \omega_{ib} \rho_c
\ee
\be
\delta \omega_i^{a} = -\partial_i \tau^{a} - \epsilon^{abc} \omega_{ib}
\tau_c \ee
Where the $\tau$ are equivalent to Lorentz transformations, whereas the
$\rho$ transformations are (on shell) equivalent to diffeomorphisms
(plus Lorentz transformations).
The system of constraints implies that the phase space consists of all
flat connections (modulo gauge transformations).Witten has shown that
(when the cosmological constant is zero, as in the action above), one
should better regard the connections $\omega$ themselves as
coordinates, and, besides, that this is a renormalizable theory, by
expanding around the "unbroken" state, $\omega = e = 0$, and imposing
the gauge condition
\be
D_{0}^{i} e_i^{a} = D_{0}^{i} \omega^{a}_{i b} = 0
\ee
(where the fidutial metric $g_0$ is unrelated to $e$ and $\omega$).
Owing to the topological character of the interaction, he also argued
that the beta function should be zero.

\par
Working out the one loop partition function one gets:
\be
Z(M) = \int De D\omega exp( i S_{EH}) = \int D\omega \prod _{ijax}
\delta(F_{ij}^{a}(x))
\ee
characterizing the moduli space of all flat $ISO(2,1)$ connections,
which is given by a mathematical expression known as the Ray-Singer
torsion. Witten went even further, and interpreted an infrared
divergence stemming from the calculations as a signal of the appearance
in the theory of a classical
regime (because the only natural scale a priori for the theory is the
Planck one and any macroscopic distance is essentially divergent with
respect to it.).
\par
When the cosmological constant is non-zero, the situation is clarified
if we introduce the natural decomposition:

\be
A_i^{a\pm} = \omega_i^{a} \pm \sqrt{\lambda} e_i^a
\ee
where the group generators are given by:
\be
J_a^{\pm} = 1/2 (J_a \pm \lambda^{-1/2} P_a)
\ee
\be
[J_a^{\pm}, J_b^{\pm}] = \epsilon_{abc} J_c^{\pm}
\ee
\be
[J_a^{+},J_b^{-}] = 0
\ee

There are then two actions of the Chern-Simons type, the "standard" one
we considered at the beginning of this paragraph,
\be
S_{EH}(A) = 1/(4\sqrt{\lambda}) (S_{EH}(A^{+}) - S_{EH}(A^{-}))
\ee
and an "exotic" possibility,
\be
S_{ex} = 1/2 (S_{EH}(A^{+}) + S_{EH}(A^{-}))
\ee
In the physical situation we are interested with in the present paper,
namely, euclidean signature (and negative cosmological constant), the
most general action can be written, after rescalings, as:

\be
S = 1/\hbar S + i k/8\pi S_{ex}(\lambda=1)
\ee
(where $k \in Z$ in order for the action to be gauge invariant).
In the semiclassical regime ($\hbar \rightarrow 0$), the partition
function
\be
Z(M) = \int De D\omega \exp{-S}
\ee
will be dominated by the classical solution of largest (negative)
action, and we would have the behavior:
\be
Z \sim \exp{-1/\hbar V + 2\pi ik C_{s}}
\ee
where $V$ is the volume of the three-dimensional manifold, and $C_{s}$
is the corresponding Chern-Simons invariant.
Witten's results allow computations of some topology changing
amplitudes; for example, the amplitude for a Riemann surface $\Sigma$ to
evolve into another Riemann surface $\Sigma'$, will be given by the
integral over the exponential of minus the acion, over the solution of
the corresponding cobordism: $\partial M = \Sigma'- \Sigma$.
It was even argued that factorization was suggested by Johnson's
results; although no matter could be included if renormalizability was
to be maintained.
Incidentaly, some computations made by Carlip and de Alwis
\cite{kn:carlipdeal} suggest that in this case at least,
the sum over
wormholes is not Borel summable, and the corresponding cosmological
constant did not appear to be driven to zero.
\par
To summarize: Witten's clever ansatz allow a computation of the
partition function for three-dimensional quantum gravity, for a fixed
topology (or rather, reduces this problem to an equivalent, but
sometimes non-trivial, mathematical question, namely, the computation
of the corresponding Ray-Singer torsion).
\par
In two-dimensional quantum
gravity, as defined by string theories, it seems neccessary, however,
to perform the sum over all topologies. Can we implement this further
in the three-dimensional case as well?
\par
A neccessary first step in this direction, is the classification of all
possible three-dimensional topologies; for example, by introducing a
complete set of topological invariants. In the two dimensional
situation,
there is such a complete set (in the closed, compact case), namely, the
Euler characteristic, $\chi = 2- 2 g$. This allows for any functional
integral in the two-dimensional situation to be written in terms of the
fixed genus one as:
\be
\int Dg = \sum_{g=0}^{\infty} \int_{fixed genus} Dg
\ee
The purpose of the following sections is to determine whether something
similar can be written in the three-dimensional case.

\section{The problem of diffeomorphic equivalence in $d=3$}
In a preceding work \cite{kn:eav5}, we studied the problem of the
classification of all topologies, up to diffeomorphic equivalence, in
the physical dimension $d=4$.
\par
The remarkable result, due to Markov, is that given two four-dimensional
manifolds, there is no algorithmic way of deciding when they are
homeomorphic. This implies, in particular, that there is no a
(countable)
set of topological invariants, such that two four-manifolds are
homeomorphic if and only if thay have the same values for all the
invariants in the set.
\par
In this case there is an additional problem, namely, that not every
homeomorphism can be lifted to a diffeomorphism.
We know, in particular, owing to the work of Donaldson, that there are
manifolds (namely, those for which a certain unimodular form over the
integers $Z$, the intersection form, is even (that is, its diagonal
elements are even ), and positive definite), which can not be smoothed.
\par
In dimension $n \geq 5$, there is an invariant, the Kirby-Siebenmann
class,(\cite{kn:ks}\cite{kn:ks2})  $e(M) \in H^4(M,Z^2)$, such that if
$e(M)\neq 0$, the topological
manifold $M$ can not be given any piecewise linear structure.(cf. the
physicist-oriented discussion in \cite{kn:nash})
\par
In low dimensions, a theorem of Rado guarantees that any topological
manifold can be upgraded to a piecewise linear structure; and in
dimension 3 the same result is due to Moise.
On the other hand, another theorem of Kirby and Siebenmann guarantees
that if the dimension is $n\leq 3$, any piecewise structure can be
promoted to a differentiable one.(cf.\cite{kn:ks2}).

In order to get Markov's result, it is enough to consider one of the
simplest topological invariants, namely, the fundamental group,
$\pi_1(M)$. As this object will play a rather important place in all
our subsequent discussions, let us pause now for a minute to remind the
reader of its definition.
\par
The object of interest is the set of all closed paths in the manifold,
that is, mappings
\be
I=(0,1) \rightarrow M
\ee
such that
\be
x(1)=x(0)
\ee
The law of composition of two such loops is simply (in a somewhat
symbolic, but otherwise evident, notation):
\be
(x_1*x_2)(u)=\theta(1/2-u).x_1(2u)+\theta(u-1/2).x_2(2u-1)
\ee
, and we are going to consider two such loops as equivalent for our
present purposes, if there is a continuous deformation of one into the
other (that is, if they are homotopic), which means  that there is a
continuous function, $f$, such that
\be
f: I \times I \rightarrow M
\ee
\be
f(t,0)=f(t,1)
\ee
\be
f(0,u)=x_1(u)
\ee
\be
f(1,u)=x_2(u)
\ee
It should be stressed that the fundamental group is, in general,
non-abelian (while the higher homotopy groups,$\pi_n(M),n\geq 2$ are all
abelian).There is a close relationship (namely, the Hurewicz
isomorphism) between the homotopy (the concept we have just defined) and
the homology (which, roughly, counts the number of holes in the
manifold). The first non-vanishing homotopy and homology groups of a
given (path-connected) manifold, occur at the same dimension, and are
isomorphic.
\par
Actually, the homology is, in a well defined sense, the
abelianization of the homotopy:
\be
H_1(M,Z)\sim\pi_1(M)/ [\pi_1(M),\pi_1(M)]
\ee

It is actually possible to enumerate all possible compact, connected
closed 3-manifolds, in the following sense: one can write down a
(countably infinite) list of sets of parameters, such that each set
specifies a three-dimensional manifold, and we are guaranteeed that
every three-manifold is contained in the list.
\cite{kn:fomenko}.We have no control over which ones in the set
are diffeomorphic  to one another, so that in general there will
be infinite overcounting as well.
\par
In the four dimensional case, every (finitely presented) group can be
realized as the fundamental group of a four-dimensional manifold; this
is not true anymore in the three-dimensional case, so that Markov's
trick of putting a non-recognizable succession of groups in
correspondence with manifolds does not appply here, which means that
there is no proof that the problem of diffeomorphic equivalence in the
three-dimensional case is undecidable (but there is no proof of the
contrary either).

\par
In spite of what has been said above, the fundamental
group
is here also the origin of some trouble. We know that
all $\pi_1(M_3)$ are finitely generated, but the problem of
characterizing those groups which can be fundamental groups of some
three-dimensional manifold is highly non-trivial. It can actually be
proved that this subclass of all finitely presented groups is
algorithmically non-recognizable; while it is also known that not every
finitely generated group can be the fundamental group of some
three-dimensional manifold; the classic example being the group $Z^4$.
\par
Although Markov's proof does not work in $d=3$, for precisely this
reason, (as we have just seen), this does not mean that the homeomorphy
problem is completely
solved. There are no definitive results in this field, and the fact is
that there is no known procedure to determine when two three-dimensional
manifolds are diffeomorphic.
\subsection{Topological invariants}
Let us quickly review the uselfulness of the known topological
invariants in the three-dimensional case (cf.\cite{kn:meyerhoff}):
\begin{itemize}
\item{The fundamental group}. 

It is too complicated to be of any
practical value.
\item{The homology}. 

This is too simple. For example, it does not
distinguish between $(S^3-K)_{(1,q)}$ and $S^3 = (S^3-K)_{(1,0)}$.
(cf.next paragraph for an explanation of the notation).
\item{Euler characteristic}.

 Actually $\chi(M) = 0$ for all cosed,
compact three.-manifolds.
\item{The volume}

Strange as it may seem, the volume is a topological
invariant for a certain type of manifolds, called "hyperbolic"; where,
by Mostow's theorem, different volumes imply non-homeomorphic
manifolds.(More on this later).
The volume is, in some sense, a measure if the complexity of the
manifold and,moreover, the set of manifolds with any given volume is
finite.
It is not however, a complete invariant: for example, the
manifold  $WL_{(1,1)}$ is not homeomorphic to $WL_{(5,-1)}$, even though
they both have the same volume.
\item{The Chern-Simons and Eta invariants}

They are actually both related (cf.\cite{kn:aps}).
\be
3\eta(M) =2 C_s (M)
\ee
the $\eta$ invariant contains obviously more information than
Chern-Simons, and besides, a theorem of Meyerhoff and Ruberman
guarantees that, given any rational number in $R/Z$, there exist
hyperbolic manifolds with equal volumes whose $C_s$ invariants differ by
that rational number.
\item{Is this enough?}

Unfortunately no, because another theorem of Meyerhoff and Ruberman
\ref{kn:meyerru}
\end{itemize}
implies the existence of certain "mutations" of closed hyperbolic
manifolds, which leave the volume, the $C_s$ (mod.1), and the $\eta$
invariants unchanged, while P. Kirk has constructed explicit examples of
non-homeomorphic mutants.
\par
In some particular cases (namely, for homology 3-spheres, that is, when
$H_1(M)=0$), there are other invariants related to knots, which will be
brielfly commented upon later on. Let us stress, however, that not even
Poincare's conjecture (that every homology 3-sphere is actually a $S^3$)
has been stablished in the three.-dimensional case.(In the
four-dimensional case this has been done in a classic work of
Freedman:\cite{kn:freedman})

\section{The geometrization conjecture}
There is a very general form of characterizing a three-dimensional
manifold (cf. \cite{kn:lickorish}). Every closed 3-manifold can be
obtained by a procedure called Dehn surgery along some link whose
complement is hyperbolic, starting from the three-sphere, $S^3$.
(a link is, by definition, a one-dimensional compact submanifold in a
three-dimensional manifold, M). The intuitive concept of what Thurston
calls Dehn surgery is quite simple: just remove a regular neighborhood
of the link K, and glue it back after some new identification. In this
way we can construct, for example, new manifolds, $(S^3 - K) _{p,q}$, by
removing a solid torus as a regular neighborhood of a closed knot $K$,
and gluing it back after having performed $p$ $2\pi$ rotations aroung
the homology cycle $A$, and $q$ $2\pi$ rotations around the other
homology cycle, $b$. The behavior of $M_{p,q}$ as $(p,q)$ get large is
well understood,; they are called cusped hyperbolic three-manifolds, and
are well described by a cartesian product of a torus times a
half-interval $[0,\infty)$ affixed to the "belly" of the 3-manifold.
\par
William Thurston
(\cite{kn:thurston1}\cite{kn:thurston2}\cite{kn:meyerhoff}), has
put forward a conjecture, known as the geometrization conjecture, and
has as well over the years produced an impressive amount of evidence in
its support; althogh neither he nor anybody else has succeeded in
proving it for the time being.
\par
The basic idea is very simple, and stems from the uniformization theorem
of Poincare, in which every two-dimensional Riemann surface is proved to
be conformally equivalent to either the sphere $S^2$ (for genus zero),
or to the quotient of the two-dimensional plane with a lattice,
$C/\Gamma$  (for genus one), or else to the quotient of Siegel's upper
half plane with a fuchsian group of the second kind  $H/G$ (for higher
genus).
\par
In two dimensions, besides, there is a constructive procedure,
(the sewing) for gluing some elementary structures (the pants, or
sphere with three-parametrized boundaries, P(0,3), and the cylinder, or
sphere with two parametrized boundaries, P(0,2) ) and obtain the generic
Riemann surface (cf, for example, \cite{kn:lag9}).

The geometrization conjecture just states that the interior of every
compact three-manifold has a canonical decomposition into pieces which
have geometric structures.
\par
This decomposition proceeds in two stages: one first performs a "prime
decomposition", by cutting along two-spheres embedded in $M_3$, so that
they separate the manifold into two parts, neither of which is a 3-ball,
and then gluing 3-balls to the resulting boundary components, thus
obtaining closed 3-manifolds which are simpler; and the second stage
involves cutting along tori in an adequate manner.
\par
And what is a geometric structure?
We shall demand, first of all, that it admits a complete, locally
homogeneous metric (That is, that for all points $x,y\in M$,there should
exist isometric neighborhoods, $U_x$ and $U_y$).
If $M$ is complete, and the space $X$ is simply-connected, then $M =
X/\Gamma$, where $\Gamma$ is a discrete subgroup of the isometry group
of $X$, $G$, without fixed points. This concept clearly generalizes the
three two-dimensional simply connected Riemann surfaces.
\par
One can show that the are precisely eight such homeogeneous spaces,
$(X,G)$, needed for a geometric characterization of 3-manifolds. We can
uniquely characterize them by demanding that $X$ be simply-connected;
and, besides,that $G$ is a group of diffeomorphisms of $X$ such that the
stabilizer of an arbitrary point $x\in X$ is a compact subgroup of $G$.
The group $G$ itself can be proven to be unimodular, which implies, in
particular, that there exists a measure right as well as left invariant.
We shall actually demand that $G$ is a maximal group of homeomorphisms
of $X$ with compact stabilizers.
\par
Let us briefly list the eight geometries (cf. \cite{kn:scott}):
\subsection{The eight three-dimensional Geometries}
\begin{description}
\item[Spherical Geometry] Here $X = S^3$ and $G=SO(4)$. The identity
component of the stabilizers at x, $G_x =SO(3)$. All three dimensional
spherical manifolds have been classified.
\item[Euclidean Geometry] Here $X = R^3$ and $G =R^3 \times SO(3)$. The
stabilizer is $G_x = SO(3)$. There are only ten non-homeomorphic
3-dimensional euclidean manifolds (of which 6 are orientable). All of
them are, moreover finitely covered by the three-torus, $T^3$.
\item[Hyperbolic Geometry] Now $X$ is the hyperbolic three-space, $X=
(x,y,z, z\geq 0)$, and $G = PSL(2,C)$. The stabilizer is $G_x=SO(3)$.
According to Beltrami's half space model, a good metric is
\be
ds^2 = 1/z^2 (dx^2 + dy^2 + dz^2)
\ee
and the points $ z= 0$ are called points at infinity.
A concrete example is the Seifert-Weber dodecahedral space. To construct
it, we identify opposite faces, after a $3/10. 2\pi$ clokwise twist. The
30 edges then reduce to 6, after identification; the 20 vertices reduce
to only 1, and the 12 faces collapse to 6.The Euler characteristic is
$\chi=V-E+F-C^{\prime}=1-6+6-1=0$, so that this construction represents
a compact, closed three-dimensional space. This procedure goes through
both in euclidean and in hyperbolic space.
This case is, in some sense, the generic one.
\item[The two-sphere cross the euclidean line].$X=S^2 \times E^1=G$.
The stabilizer is $G_x =SO(2)$.

There are only two non-homeomorphic examples of compact manifolds with
this geometry. It is however, an important case from the physical point
of view, because it represents a two-manifold of spherical topology
evolving in time.
\item[Hyperbolic two-space cross the euclidean line].$X=H^2 \times
E^1$,$G=IS(H^2) \times IS(E^1)$. The stabilizer is

$G_x=SO(2)$. Every manifold modelled on this geometry is finitely
covered by the product of a surface and a circle. This case is also very
important physically, because it represents a two-manifold homeomorphic
to a generic Riemann surface, evolving in time.
\item[Universal covering of the special group]$X =
\overline{SL(2,R)}$,$G = R \times
\overline{ISOM(H^2)}$.
The universal covering of $SL(2,R)$. The stabilizer is $G_x =SO(2)$. The
space of unit tangent vectors to any hyperbolic surface is an example of
a manifold with this geometry.
\item[Heisemberg group] Here $X= Nil$,$G =H \times_{s} S^1$.
The stabilizer is $SO(2)$.
Any oriented circle bundle over a 2-torus, $T^2$, which is not the
three-torus, $T^3$, has this kind of geometric structure.
A concrete realization consists of the upper triangular matrices with
unit diagonal,
\be
\left(\begin{array}{ccc}
1&x&z\\0&1&y\\0&0&1
\end{array} \right)
\ee
which can be interpreted as $R^3 =(x,y,z)$, with the composition law:
\be
(x,y,z).(x^{\prime},y^{\prime},z^{\prime})
=(x+x^{\prime},y+y^{\prime},z+z^{\prime}+xy^{\prime})
\ee
and the line element given by:
\be
ds^2 = dx^2 + dy^2 +(dz -  x dy)^2
\ee
\item[Sol]. Here $X=Sol$ (a soluble group), and $G$ is an extension of
$X$ by $Z_2$. The stabilizer is now trivial $G_x=1$.
Any torus bundle over $S^1$  whose monodromy is a linear map with
distinct, real eigenvalues has a geometric structure of this form.
Actually, here also the space can be taken as $R^3 =(x,y,z)$, with the
composition law:
\be
(x,y,z).(x^{\prime},y^{\prime},z^{\prime})
=(x+\exp{-z}x^{\prime},y+\exp{-z}y^{\prime},z+z^{\prime})
\ee
and the metric given by:
\be
ds^2 = \exp{2z} dx^2 + \exp{-2z} dy^2 + dz^2
\ee
\end{description}

\par
Starting from a description of the (supposedly hiperbolic) manifold
using Dehn surgery along
a link , to compute the geometrical decomposition one must first of all,
calculate representations of $\pi_1(M) \rightarrow PSL(2,C)$ (the group
of isometries of $H^3$). Then one has to check whether a given
representation is discrete and faithful, and finally one ends up with a
hyperbolic manifold whose fundamental group is $\pi_1(M)$. The procedure
makes clear the fact that we have not control ovel homeomorphic
equivalence in this process.
\par
We have mentioned earlier on Mostow's rigidity theorem. Stated somewhat
more precisely, it says that if two hyperbolic manifolds of finite
volume have isomorphic fundamental groups, they must necessarily be
isometric to each other. This clearly implies that if a closed,
orientable three-dimensional manifold possesses a hyperbolic structure,
then this structure is unique (up to isometry).
\par
In the non-compact case there is a curious non-rigidity theorem due to
Thurston, which states that if $M=H^3/\Gamma$ is orientable and
non-compact, with finite volume, $V$, then there is a succesion of
manifolds, $M_j =H^3/h_j(\Gamma)$, with volumes strictly smaller than
$V$, $V_j< V$, and such that $lim_{j\rightarrow \infty}M_j = M$, in some
precise sense.
\par
These structures can sometimes be characterized as (Seifert) bundles
$\eta$ over orbifolds $Y$.
\footnote{This means that there exists a decomposition of $M$ into disjoint
circles (called fibres) such that each circle has a neighborhood in $M$
which is a union of fibres and is isomorphic to a fibred solid torus or
Klein bottle}
\par
Classifying them according to the Euler numbers of the orbifold,
$\chi(Y)$, and the Euler number of the bundle,$e(\eta)$, we get:
\be
\begin{array}{cccc}
*&\chi(Y)>0&\chi(Y)=0&\chi(Y)<0\\e(\eta)=0&S^2\times
R&E^3&H^2\times R\\e(\eta)\neq 0&S^3&Nil&\overline{SL_2} \end{array}
\ee

\section{New invariants}

There has been recent progress in defining new invariants, which,
are only useful, however, in some simplifying situations.
Floer, in particular, defines invariants of homology $3$-spheres
(that is, oriented, closed, $3$-dimensional smooth manifolds such
that  $H_1 (M,Z)=0$). Other (finer) invariants, defined by Casson, are,
in a sense, one half the Euler characteristic of this homology
groups.
To be specific, in order to define the Casson invariants, one has
to study the representations of the fundamental group into $SU(2)$,
and introduce a precise way of counting them. One then has:
\be
\lambda(M_3) = 1/2 ( Irreps:\pi_1(M_3) \rightarrow SU(2) )
\ee
The precise form of the relationship with Floer homology is then,
\be
\lambda (M_3) = \sum_{i=0}^{7}(-1)^{i} dim F_{i}^{+}(M_3)
\ee
The celebrated Jones polynomials, originally devised to characterize
knots and links, can be generalized as well to some three-dimensional
compact manifolds.
\par
The intuitive way of constructing them
(cf.\cite{kn:kaufman};\cite{kn:reshetikin};\cite{kn:turaev}
\cite{kn:turaev2})
is to start
with the known representation of a three-dimensional manifold as the
result of performing surgery on a link on a three-dimensional sphere
$S^3$.
The actual computations performed by Rehetikin and Turaev, and later on,
by
Turaev and Viro, rely on quantum group techniques to build up appropiate
averages of link polynomials. The two invariants are deeply related,
and, for example, if $M_q$ denotes the Turaev-Viro invariant, and
$Z_q(M)$ denotes the Reshetikin-Turaev invariant for $SU(2)_q$, one has
the relationship
\be
M_q = Z_q(M) \overline{Z_q(M)}
\ee

Many new relations among all these invariants are constantly being
unveiled, and it is a topic of current research among mathematicians -
and also among physicists, thanks to the topological quantum
field theories of Atiyah and Witten . (cf. Atiyah's lucid reviews
in \cite{kn:atiyah3}\cite{kn:atiyah6}\cite{kn:atiyah7}) .As he likes to
put it, one of the most important

open questions in three-dimensional topology is the relationship between
this type of invariants and the geometrization programme of the fourth
paragraph.

\section{Some provisional conclusions}

Let us imagine that Thurston's conjecture is true. This means that we
can represent every compact, closed, three-dimensional manifold as
\be
M_3 = \bigcup _{i=1}^{\infty} G_{n_i} /\Gamma_{n_i}
\ee
where $n_{i}\in (1,...8)$ represents one of the eight geometries, and
$\Gamma$ is a subgroup pf the isometry group of the corresponding
geometry.
The glueing in the preceding formula will be represented by a
complicated (although computable in principle) set of moduli.
\par
If each glueing is characterized  by a different coupling constant (in
the same way that in string theory the gluing of the pants and the
cylinder is characterized by the single coupling constant $\kappa$,
later to be identified with Newton's constant)
then we would have $ 28$ different coupling constants $\kappa_i$, whose
physical interpretation is perhaps possible without solving all problems
of definition of the path integral.
\par
It is also possible that, in the same way as in the open string case,
in which two apparently unrelated coupling constants turn out to be the
same due to consistenvy requirementes, that here also not all of the
$\kappa_i$ are independent, and consistenmcy forces to implement
nontrivial relations betwween them.
\par
The (symbolic) form of the measure would then be:
\be
Dg \sim \sum_{combinations} D(Isometry Groups) D(Moduli)
\ee
When one thinks that even in the two-dimensional case, the integration
region is not explicitly known for genus $g \geq 4$, (and, besides,
the action of the mapping class group on  the Fenchel-Nielsen
coordinates is very complicated, so that one has in practice to use
other coordinates on moduli space, much less naturally related to the
pants decomposition), one can easily
realize that both measures in the preceding formula are very difficult
mathematical problems.
\par
Perhaps this can be done, however, with a bit of luck. But then, we have
to face the main problem, and that is that nowhere in the preceding
discussion we have taken into account that we have to sum in the path
integral over non-homeomorphic manifolds only. This means that there
will be infinite overcounting, and it seems impossible to overcome this
problem unless some numerable set of topological invariants is devised
which characterize completely a manifold up to diffeomorphisms.
\par
Please note that this problem will still be there, even though the
fundamental theory of gravity turns out to be topological (at least in
the absence of matter). And this is because, although in this case one
does not need to solve the homeomorphy problem, one needs, however, to
be able at least to "count" and enumerate all topologies. Besides, the
topological phase has to be broken at long distances, and there, in the
description of the symmetry breaking, one has to face again all the
problems of the homeomorphic equivalence.

\newpage

\section*{Acknowledgements}
I am grateful to Luis Alvarez-Gaume, JLF. Barbon, J.M. Gracia-Bondia, P.
Hajicek, Y. Lozano, MAR. Osorio and MA. Vazquez-Mozo for useful
discussions and suggestions.


\end{document}